\documentclass[twocolumn,showpacs,preprintnumbers,prd,amsmath,amssymb]{revtex4}
\usepackage[dvips]{graphicx}
\usepackage{amsthm}
\usepackage{amstext}
\usepackage{amsfonts}
\usepackage{hyperref}
\newcommand{\p}{\partial}

\newcommand{\dd}{{\rm d}}

\begin{document}

\title[Simultaneity and generalized connections in general relativity]{Simultaneity and generalized connections in general relativity } 

\thanks{Published version: E. Minguzzi, Class. Quantum Grav. 20 (2003)
2443-2456. Classical and Quantum Gravity copyright \copyright
(2003) IOP Publishing Ltd,
\href{http://www.iop.org/EJ/journal/CQG}{http://www.iop.org/EJ/journal/CQG}}
\author{E. Minguzzi}
 \affiliation{Dipartimento di Fisica dell'Universit\`a di Milano-Bicocca,\\ Piazza della Scienza 3, 20126 Milano, Italy\\ minguzzi@mib.infn.it }

\begin{abstract}
Stationary extended frames in general relativity are considered.
The requirement of stationarity allows  to treat the spacetime as
a principal fiber bundle over the one-dimensional group of time
translations. Over this bundle a connection form  establishes the
simultaneity between neighboring events accordingly with the
Einstein synchronization convention. The mathematics involved is
that of gauge theories where a gauge choice is interpreted as  a
global simultaneity convention. Then simultaneity in
non-stationary frames is investigated:  it turns to be described
by a gauge theory in a fiber bundle without structure group, the
curvature being given by the Fr\"olicher-Nijenhuis bracket of the
connection. The Bianchi identity of this gauge theory is a
differential relation between the vorticity field and the
acceleration field. In order for the simultaneity connection to be
principal, a necessary and sufficient condition on the 4-velocity
of the observers is given.
\end{abstract}

\maketitle


\section{Introduction}
Since the early days of  special relativity,   the isotropy of the
speed of light was considered as conventional, and related to the
synchronization convention of distant clocks.  The discussion was
revived in the last decades by a controversial experimental test
of the isotropy of the speed of light \cite{will92,anderson98} and
by arguments which showed the privileged status of the Einstein
synchronization convention over other conventions
\cite{malament77,anderson98}. On the other side, some authors
\cite{anderson98,minguzzi02} reconsidered those results proving
the  independence of the laws of physics from the synchronization
convention (simultaneity convention) adopted, and the existence of
a set of conventions alternative to the isotropic one. Along these
lines, it was shown that the causal structure of
 Minkowski spacetime follows solely from the constancy of the speed of light
over closed paths \cite{minguzzi02,minguzzi02d}. This   provided
new confidence on the local Minkowski nature of the spacetime
manifold, and suggested how to make statements without relying on
synchronization conventions. The strategy was    to connect
convention-free quantities (i.e. independent of the
synchronization convention), like the two-way speed of light, to
convention-free quantities, like the causal structure. This
convention-free approach will prove to be useful  in more general
situations.

Leaving inertial frames in special relativity  things become far
more clear. In general it becomes impossible even to synchronize
distant clocks using the Einstein synchronization convention and
hence no question arises on its predominance over other
conventions. One cannot avoid the arbitrariness of the choice of a
global synchronization. A formalism would be welcome if it is able
to provide  quantities independent of the synchronization
convention adopted. This formalism is that of gauge theories. The
aim of this work is to fix completely the parallelism between
conventionality of simultaneity and gauge theories. The need for a
gauge formalism was already noticed \cite{anderson92,minguzzi02}.
Here, a coherent exposition in light of Ehresmann's theory of
connections \cite{kobayashi63} is given. It has the great
advantage of unifying old  and new results (see
Eq.~(\ref{bianchi}) and its interpretation as a Bianchi identity)
in a common formalism which provides new insights on the geometry
of spacetime.

We start with stationary frames in general relativity. Our
conclusions will have straightforward specializations in  usual
contexts like the rotating platform in special relativity or the
Kerr metric in general relativity. Then we study simultaneity in
non-stationary frames showing that it is still described by a
gauge theory in a generalized sense already developed by
mathematicians \cite{michor91,mangiarotti84,modugno91}. The
derivation is short but needs at least some notions of
differential geometry of principal fiber bundles
\cite{kobayashi63,wu75}. We shall use the conventions of Kobayashi
and Nomizu \cite{kobayashi63} apart from the wedge product that
here is fixed by $\alpha \wedge \beta=\alpha \otimes \beta - \beta
\otimes \alpha$. The spacetime metric has signature (+ - - -).

\section{The geometry of extended reference frames} \label{space}
In everyday experience we deal with reference frames extended over
large scales. On the Earth's surface,  physics laboratories
communicate to each other  considering themselves at rest with
respect to the same global reference frame. Such a situation is
very different from the local inertial reference frame that can be
constructed in the neighborhood of a given  free falling, non
rotating,  observer. There, locally, the usual Minkowski spacetime
is recovered and the Euclidean spatial metric (that experienced by
rods at rest with respect to the observer) is derived as the
projection of the spacetime metric over the simultaneity slice,
where the simultaneity of events is defined by the Einstein
convention. Things change when a large number of observers are
brought together to form an extended reference frame. For
instance, the Einstein simultaneity convention is unsuited for
rotating systems because clocks cannot be synchronized in that way
all around a closed path. The spatial metric can no more be
defined as the projection of the spacetime metric over the
simultaneity slice because the Einstein synchronization
convention, in general, does not work in the large and, therefore,
such a slice does not exist. Hence, the spatial metric needs  a
more general definition (see section \ref{section}).

Let $M$ be the spacetime manifold of metric $g_{\mu \nu}$, and let
$k$ be a time-like Killing vector field. We shall say that a body
is at rest with respect  to the stationary frame if its worldline
is an integral line of the Killing vector field. Its four-velocity
is (we shall use the symbol $u$ also to denote the 1-form
 $u_{\alpha}$; its meaning will be clear from the context)
\begin{equation} \label{four}
u=\frac{k}{\sqrt{k \cdot k}}.
\end{equation}
Let $G=T_{1}$ be the one-parameter group of diffeomorphisms
generated by the Killing vector field and let $t$ be the
corresponding parameter, $k=\p_{t}$. The previous definition
identifies a point of space of the stationary frame with an
integral curve of the Killing vector field, hence the "space" is
defined by
\begin{equation}
S=M/G,
\end{equation}
that is, as the quotient space of $M$ under the action of the Lie
group $G$. Here, we recognize, at least locally, all the
ingredients of a principal fiber bundle: $M$ is the principal
bundle, $G$ is the structure group and $S$ is the base. We need
only to define the right action of $g=e^{t} \in G$ on $m \in M$ as
$mg \equiv \phi_{t}(m)$ where
 $\phi_{t}$ sends an event to its evolution after a Killing time
$t$. The fiber $F_{s}$  over $s \in S$ is given by the integral
line identified with $s$, and we shall refer to it also as the
worldline of $s$. For definiteness it is assumed that the
projection $\pi: M \to S$ of the bundle onto the base is
differentiable. Finally, the principal bundle is trivial because
its structure group,  the group of translations in one-dimension,
 is contractible \cite{steenrod70}.

 The systematic construction of the gauge theory  would be complete
 if we could provide a 1-form $\omega$ (the connection)  on $M$, with the following
 properties \cite{kobayashi63}
 \begin{eqnarray}
&(a)& \qquad  \omega(k)=1, \label{onormalization} \\
&(b)& \qquad \phi_{t}^{*} \omega = \omega. \label{hinvariance}
 \end{eqnarray}
where $\phi_{t}^{*}$ is the pullback of $\phi_{t}$. This is
accomplished by the choice
\begin{equation} \label{connessione}
\omega=\frac{k_{\mu}\dd x^{\mu}}{k \cdot k}.
\end{equation}
Statement (a) follows trivially from substitution  and statement
(b) from the invariance of (\ref{connessione}) under translations:
$ L_{k}k=L_{k}g=0$. In a given event $m$, the tangent space
$T_{m}$
 splits in two parts, the vertical space $V_{m}$ generated by the
Killing vector field, and the horizontal space $H_{m}$ orthogonal
to it: $\omega(X)=0 \leftrightarrow X \in H$.

Let us investigate more closely the physical meaning of this gauge
theory.  An observer at rest in the stationary frame can define,
in his neighborhood, a simultaneity slice in compliance with the
Einstein synchronization convention.
\begin{figure}[!ht]
\centering
\includegraphics[width=6cm]{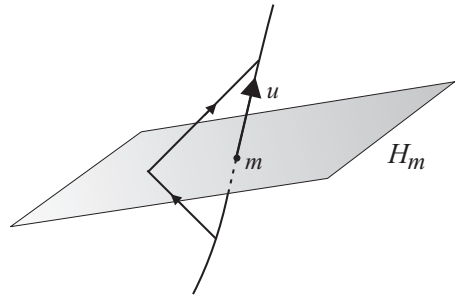}
\caption{The horizontal plane and its relation with Einstein's
simultaneity.} \label{horizontal}
\end{figure}
The coordinates obtained following the Einstein convention, even
for events far away from the worldline, are known as
M\"arzke-Wheeler coordinates \cite{marzke64,pauri00}. Here we are
interested only in  the trivial result that the worldline is
indeed orthogonal to the simultaneity slices. This result should
be expected because it is a local statement that holds in special
relativity. At least locally this means, as it was emphasized by
Robb \cite{robb14,anderson98}, that the Einstein convention
corresponds to taking as simultaneous with $m \in M$ those events
in a neighborhood of $m$ that lie in the exponential map of the
plane perpendicular to $u$. From this we see that the horizontal
space $H_{m}$ identifies those events which are simultaneous with
$m$ with respect to an observer at rest in the stationary frame
and placed in $s=\pi(m)$. In the following we shall refer to
$\omega$ as the Poincar\'e connection, because Poincar\'e
\cite{poincare04a,poincare04b} was the first who defined the
synchronization convention that, later, was named after Einstein.

Let $\tau(\lambda)$ be a curve on $S$. The observers, all along
$\tau$,  synchronize the neighboring clocks using the Einstein
synchronization convention. Given an event $\tau^{*}(0)$,
$\pi(\tau^{*}(0))=\tau(0)$,   they find a curve
$\tau^{*}(\lambda)$ of simultaneous events. The procedure that we
have sketched  is the natural one that should be followed in
synchronizing clocks, since it is simply obtained by using the
prescriptions of the Einstein convention in each point of space.
The reader should distinguish between this procedure and the one
that leads to the M\"arzke-Wheeler coordinates since in the latter
case an observer is privileged. In order to avoid confusions, in
what follows we shall refer to the Einstein synchronization
convention as a local procedure to be applied in each point of
space. As we shall see this procedure cannot be applied in every
circumstance \cite{anderson98}. In gauge theory the curve
$\tau^{*}$ is called the horizontal lift of $\tau$;  it is the
only curve which starts from $\tau^{*}(0)$, has horizontal tangent
vectors, and projects onto $\tau$.
\begin{figure}[!ht]
\centering
\includegraphics{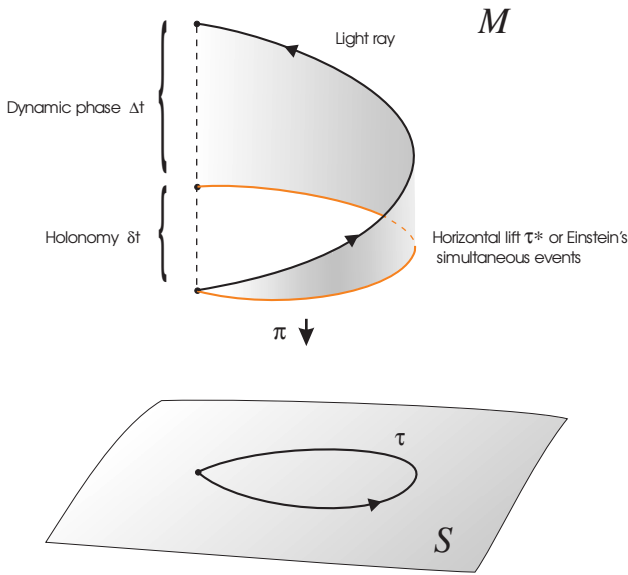}
\caption{The horizontal lift and its relation with the pointwise
Einstein synchronization.} \label{holonomy}
\end{figure}

 Before we start
studying the curvature and its implications, we recall some
notions regarding local observers. An observer is identified by a
tetrad of orthonormal vectors $\{ u, e_{1}, e_{2}, e_{3} \}$.  A
picture of $e_i$ may be the following: imagine the local
laboratory of the observer to be a cube. The observer, at rest in
the cube, stands by one of the corners and the normal vectors
$e_{i}$ point towards the next three corners.  Since $e_{i}$
points always towards the same corner it must satisfy
\begin{equation}
L_{k}\,e_{i}=0.
\end{equation}
 Let $h^{\mu}_{\nu}=\delta^{\mu}_{\nu}-u^{\mu}u_{\nu}$ be the projector
on the horizontal space of the observer. The evolution of the
components of $e_{i}$ with respect to a Fermi transported tetrad
(non-rotating reference frame) $\hat{e}_{\mu}$ is  well known
\cite{hawking73}, the relevant quantities are the vorticity tensor
\begin{equation} \label{vorticity}
w_{\mu \nu}=h^{\alpha}_{\mu}h^{\beta}_{\nu} \,u_{[\alpha ;
\,\beta]}\, ,
\end{equation}
which represents the angular velocity of the triad $e_i$ with
respect to gyroscopes ($\hat{e}_{i}$), and the expansion tensor
\begin{equation}
\theta_{\mu \nu}=h^{\alpha}_{\mu}h^{\beta}_{\nu} \,u_{(\alpha ;\,
\beta)}\, ,
\end{equation}
which represents the rate of separation of neighboring points from
the worldline.  From Eq.~(\ref{four}), after some algebra, we find
that the expansion tensor vanishes thus allowing a local rigidity
interpretation. The vorticity can be rewritten
\begin{equation}
w_{\mu \nu}=\sqrt{k \cdot k} \,\,\omega_{[\mu ;
\,\nu]}=-\frac{\sqrt{k \cdot k}}{2}\, \Omega_{\mu  \nu},
\end{equation}
where we have introduced the curvature
\begin{equation}
\Omega= D\omega =(\dd \omega) h=\dd \omega.
\end{equation}
The translational invariance of spacetime implies that the
vorticity, the curvature and the norm of the Killing vector field
have a vanishing Lie derivative. As a consequence the angular
velocity of an observer at rest in the stationary frame points
always in the same direction with the same norm. Its components
with respect to $e_{i}$ do not change in time.

Gauge theory tells us that if $S$ is paracompact and simply
connected the curvature  vanishes if and only if the horizontal
lifts of closed curves in $S$ are closed in $M$. In order to make
use of the Einstein  convention the horizontal lift should be
closed, otherwise there would be two  simultaneous events  placed
in the same time-like worldline. In that case, the Einstein
convention would be of little use. If the topological requirements
are satisfied: the Einstein synchronization convention is
applicable if and only if, in every point of the stationary frame,
observers at rest do not rotate \cite{godel49}. Moreover,  if the
vorticity is different from zero in a point $s$ of space, an
observer at rest in that point  feels inertial forces because its
local reference frame rotates with respect to Fermi transported
gyroscopes. These observations assure  that the rotating platform
in special relativity is indeed quite representative of the
general interrelationship between  the Einstein convention,
inertial forces, and rotation.

Let us study the simple case $\Omega=0$. Gauge theory tells us
that, when the curvature vanishes, there exists a foliation of the
principal bundle in three-dimensional hypersurfaces with
horizontal tangent spaces. They are the hypersurfaces of
simultaneity accordingly with the Einstein convention. Those
hypersurfaces are orthogonal to the Killing vector field because
their tangent spaces are also orthogonal to it. The existence of
hypersurfaces orthogonal to a vector field is the object of
Frobenius theorem: the necessary and sufficient condition for the
vector field $v^{\mu}$ to be hypersurface orthogonal is
$v_{[\alpha} \nabla_{\beta} v_{\gamma]}=0$. This condition is
indeed coherent with $\Omega=0$ because the equations
\begin{eqnarray}
\Omega&=&0, \\
k \wedge \dd k &=&0,
\end{eqnarray}
are algebraically the same.

If $\Omega \neq 0$ the Einstein convention is no longer
applicable. Here gauge theory suggests a standard procedure to
extend coordinates $\{x^{i}\}$ on $S$ to the whole principal
bundle. First  choose a   section $\sigma: S \rightarrow M$ and
then associate to the event $\sigma(x^{i})e^{t}$ the coordinates
$\{x^{i}, t\}$. This procedure depends on the section (gauge)
chosen and, from the physical side, corresponds to taking the
events on the hypersurface $\sigma(S)$ as simultaneous. A number
of different simultaneity conventions can be adopted, each one in
correspondence with a particular section. In geometry the change
of section is exactly what in physics is  a gauge transformation.
Hence we pass from one convention to another via gauge
transformations. Usually in physics one works with the potential,
that is with the pullback $A_{\sigma }(x^{i})=\sigma^{*}\omega$.
In the coordinates $\{x^{i}, t\}$, the Poincar\'e connection
becomes
\begin{equation} \label{connesione2}
\omega= \dd t + A_{\sigma}(x^{i}) .
\end{equation}
It is invariant under  the change of section (gauge
transformation)
\begin{eqnarray}
  t' &=& t-\alpha(x^{i}), \\
A_{\sigma'} &=& A_{\sigma}+\dd \alpha(x^{i}).
\end{eqnarray}
It is also convenient to introduce the field strength
$F(x^{i})=\sigma^{*} \Omega=\dd A_{\sigma}$. Let us come to a
frequent question in gauge theories. If $\tau(\lambda): [0,1]
\rightarrow S $ is a closed curve, how much is $\delta t$ such
that
 $\tau^{*}(1)=\tau^{*}(0)e^{\delta t}$? In order to answer this question we
 have to recall that $\tau^{*}$ has horizontal tangent vectors.
 Integrating $ \omega$ over $\tau^{*}$ we find
 \begin{equation} \label{deltat}
\delta t=-\oint_{\tau} A_{\sigma}.
 \end{equation}
The physical interpretation of this result is connected with the
Sagnac effect \cite{sagnac13,anandan81}.

\section{Space metric and  Sagnac effect} \label{section}
 Let us define
over $S$ a metric through the formula
\cite{moller62,gron77,landau62,weber97}
\begin{equation}
\dd l^2= -h_{i j} \dd x^{i} \dd x^{j} \qquad i,j=1,2,3
\end{equation}
in the coordinates $\{ x^{i}, t\}$  introduced above. From a
geometrical perspective this formula for the space metric $\dd
l^2$ can be rewritten
\begin{equation} \label{space metric}
\dd l^2(X, Y)=-g(X^{*}, Y^{*}) \, ,
\end{equation}
where $X^{*}$, $Y^{*}$, are the horizontal lifts of the space
vectors $X$, $Y$. The vectors $X^{*}$, $Y^{*}$ belong to the
horizontal space of the observer which passes through their
origin, and that horizontal space is nothing but the local space
of an inertial observer. Hence, from a physical point of view the
previous formula says that the space metric is induced from the
measurements made by local inertial observers. This statement
relies on the  fact that, for an inertial observer, the spacetime
metric serves also to define the space metric through the formula
$\eta(X^{*}, Y^{*})$ where $X^{*}$, $Y^{*}$ belong to its  space.
However it is an assumption that for accelerating or rotating
observers, space measurements should coincide with those performed
by inertial observers. This postulate referred also as the
"surrogate frames postulate" \cite{klauber98} is waiting an
experimental verification. Actually, in our stationary frame,
observers at rest in the frame are not necessarily inertial. To
claim that equation (\ref{space metric}) describes accurately the
space metric as experimented by observers at rest in the frame,
 would go beyond our knowledge. A number of alternatives, especially
for the rotating platform have been suggested
\cite{strauss74,klauber98}. A similar debate raised in the contest
of Kaluza-Klein extensions of gravity where two different
interpretations of the five dimensional metric were natural: those
of Weyl and Pauli \cite{overduin97}. This is not accidental as the
mathematics involved in the spacetime splitting in time plus space
is that of Kaluza -Klein theories. Indeed the spacetime metric can
be rewritten
\begin{equation} \label{kaluzam}
g(A,B)=(k \cdot k) \omega(A) \omega(B)-\dd l^2(\pi_{*} A,
\pi_{*}B)\, ,
\end{equation}
which is a Kaluza-Klein metric in the spacetime manifold. Another
useful choice for the space metric is the optical metric
\cite{abramowicz88}, $\dd \bar{l}=\dd l/(k \cdot k)$.  Indeed the
geodesics of light, and hence their projections, are invariant
under conformal transformations of the spacetime metric $g$. In
the study of light propagation we can reduce the spacetime metric
to the form, $\dd \bar{s}^{2}=\omega^{2}-\dd \bar{l}^{2}$ where no
conformal factor appears.  Now,  the projection of a geodesic of a
Kaluza-Klein metric with unitary scalar field, $k \cdot k=1$,
satisfies a straightforward generalization of the Lorentz law. In
the null case it reads
\begin{equation}
\bar{\nabla}_{v}v^{i}= F^{i}_{j}v^{j}\, ,
\end{equation}
where $v^{i}=\dd x^{i}/\dd \bar{l}$ is the tangent vector of the
projection and $\bar{\nabla}$ is the covariant derivative with
respect to the optical Levi-Civita connection. Hence, at least
light rays, satisfy a Lorentz like equation in the optical metric
background.
 Here we use the optical metric  as
a tool for finding the dynamic phase(see figure \ref{holonomy}).

Let us consider two light beams that leave $\tau^*(0)$  following
the path $\tau(\lambda)$, one in the positive direction and the
other in the negative direction. Figure \ref{holonomy} shows one
of the beam and displays some relevant quantities like the dynamic
phase $\Delta t$ and the holonomy $\delta t$.
 As seen from the local inertial frames along $\tau$, with
respect to their local Einstein synchronized times, the two light
beams have the same speed $c=1$. Let $l$ be the natural parameter
 of $\tau$. The dynamic phase is
\begin{equation}
\Delta t=\int_{\tau^{*}} \dd t= \int_{\tau^{*}} \frac{\dd
\tau}{\sqrt{k \cdot k}}= \oint_{\tau} \frac{\dd l}{c\sqrt{k(l)
\cdot k(l)}}=\bar{L}/c \, .
\end{equation}
Where $\bar{L}$ is the total optical length of the path $\tau$.
Coming back to the Sagnac effect, the dynamic phase does not
depend on the direction followed by the light beam. On the
contrary, the holonomy changes sign; as a consequence the
difference of the arrival times of the two beams is twice
Eq.~(\ref{deltat}), and taking into account the time dilation,
that is the difference between Killing time and proper time, the
Sagnac effect becomes
\begin{equation}
\delta \tau=-2\sqrt{k \cdot k}\oint_{\tau}A_{\sigma} \, .
\end{equation}
This equation is conveniently written in the coordinates $\{x^{i},
x^{0}=t\}$. Let $\Sigma$ be a two-dimensional surface such that
$\tau=\p \Sigma$. From $k=\p_{t}$ we have $k_{\mu}=g_{\mu \nu}
k^{\nu}=g_{\mu 0}$ and $A_{\sigma i}=g_{0 i}/g_{0 0}$, hence
\cite{ashtekar75}
\begin{equation} \label{sagnac3}
\delta \tau=-2(g_{0 0})^{\frac{1}{2}}\oint_{\tau}\frac{g_{0
i}}{g_{0 0}} \,\dd x^{i}=2 (g_{0
0})^{\frac{1}{2}}\int_{\Sigma}\frac{w_{i j}}{(g_{0
0})^{\frac{1}{2}}}\, \dd x^{i} \wedge \dd x^{j} \, .
\end{equation}
Notice that the  worldline of a point at rest in the stationary
frame has equation, $x^{i}=const.$. This condition, and the
independence of $g_{\mu \nu}$ of time,  selects those systems of
coordinates for which the previous equation holds.

We have completed our exposition of the gauge property of
simultaneity in stationary frames. The reader familiar with gauge
theories can easily recognize the formal similarity with
electrodynamics. Both theories have  the mathematical structure of
a gauge theory over a one-dimensional group. The only relevant
difference is that in the stationary frame the structure group is
contractible (otherwise there would be closed time-like curves),
whereas in electrodynamics, $U(1)$, is not. This implies that, in
electrodynamics, we can have non trivial principal bundles like
monopoles, whereas this is not possible in the present context.
Other analogies, like that between the Sagnac effect and the
Aharonov-Bohm effect \cite{sakurai80}, or like
 that between magnetic forces and Coriolis forces \cite{coisson73,semon81}, become self
evident in light of the gauge interpretation.

\begin{table}
    \begin{center}
        \begin{tabular}{|c|c|}
            \hline
            & \\
             {\bf Gauge theory}  &  {\bf Simultaneity}  \\
             & \\
            \hline\hline
            Fiber bundle  & Spacetime manifold \\
            $P$ & $M$ \\
            \hline
            {\em Structure group}  & {\em Group of time translations}  \\
            $G$ & $T_1$ \\
            \hline
            Base & Space of worldlines \\
            & $S$  \\
            \hline
            Fiber & Worldline of an observer \\
            & at rest in the frame \\
            \hline
            Vertical space at  & Space spanned by the four  \\
             the point $p$ & velocity of an observer at $p$ \\
            \hline
            Horizontal space & Local Einstein's \\
            & simultaneous events \\
            \hline
            Horizontal lift & Pointwise Einstein's \\
            & synchronization  \\
            \hline
            Connection & Poincar\'e connection $P$ \\
            \hline
            Choice of the section & Choice of the simultaneity \\
            & convention \\
            \hline
            Gauge transformation  & Change of  the simultaneity \\
            & convention \\
            \hline
            Curvature &  Angular velocity of observers \\
            &  at rest in the frame \\
            \hline
            Bianchi identity &  $Dw+a \wedge w=0$ \\
            \hline
            {\em Holonomy} &   {\em Sagnac effect}  \\
            \hline
            {\em Dynamic phase} & {\em Optical round trip time} \\  & {\em in absence of
            holonomy} \\ & $\bar{L}/c$  \\
            \hline
            {\em Kaluza-Klein metric} & {\em Spacetime metric}  \\
            \hline
        \end{tabular}
    \end{center}
    \caption{Gauge interpretation of simultaneity. The correspondences in italic  hold  in stationary frames but do not hold   in general. }
    \label{table}
\end{table}

\section{Simultaneity in non-stationary frames}
So far we have considered stationary frames. In general, however,
the trajectories of the observers which define the frame, are not
 generated by a Killing vector field. This does not imply that the
simultaneity is no more described by a gauge theory, but only that
there is no  connection preserving structure group. We can still
define the space $S$ as the manifold of the observer worldlines.
Then, discarding some pathological choices for the trajectories,
at least in an open set, the space $S$ will be a differentiable
manifold and there will be a differentiable projection $\pi: M
\rightarrow S$. Hence, again, $M$ can be considered as a fiber
bundle over the space $S$.

It should be noticed that over this bundle one can define the
action of the one-parameter group of diffeomorphisms
$\phi^{u}_{\tau}$ generated by the four-velocity field $u(x)$.
With this right action $M$ can be considered, at least locally, as
a principal bundle. However, in general, such a structure group
does not preserve the simultaneity connection. It is, then, more
convenient to treat $M$ as a fiber bundle without structure group
since the beginning.

The theory of connections, i.e. gauge theory, was developed by
mathematicians in this general context \cite{michor91}.
Generalized connections appeared also in physics  in some works on
Kaluza-Klein theory and general relativity \cite{cho92,Yoon99}. In
those papers it was shown that most of the usual gauge theory has
an analogue in the new setting even from the dynamical side.
However, they did not take advantage of the coordinate-independent
approach developed by mathematicians.

Let us introduce the generalized connection in a form useful for
our purposes. A connection is given by a distribution of planes
$H_{x}$, that is, by a differentiable assignment, in each point of
the bundle, of a plane that we shall call horizontal. The plane,
together with the tangent space of the fiber  spans the tangent
space at $x$. The relevant difference with the usual gauge theory
is the absence of a structure group which sends the fiber to
itself and horizontal planes to horizontal planes. Many of the
features of the usual gauge theory have a generalization. It is
still possible to define the horizontal lift, the curvature and
the parallel transport. Even the Bianchi identities have a
generalization \cite{michor91}.

In our case, the fiber bundle M has a  natural connection. The
horizontal planes are given by $H_{x}=\text{ker} (u_{\alpha}(x)
\dd x^{\alpha})$ where $u^{\alpha}$ is the four-velocity of the
observer. In this way the connection maintains its original
meaning: the horizontal plane still selects, locally, those events
which are simultaneous accordantly with the Einstein
synchronization convention. Analogously, the horizontal lift
$\tau^{*}$ is obtained by a pointwise Einstein synchronization
over $\tau$. Notice that the connection $\omega=u_{\alpha} \dd
x^{\alpha}$ is defined only up to a spacetime factor, because its
ker does not change after such a multiplication. For this reason,
in the generalized gauge theory, one works with the vector valued
connection $P^{\beta}_{\alpha}=u^{\beta}u_{\alpha}$, which is the
projector of the tangent space onto the vertical subspace. The
curvature is defined through  the Fr\"olicher-Nijenhuis bracket
which acts on a pair of vector valued forms, say $K$ and $L$ of
degree $k$ and $l$ respectively, giving a vector valued form of
degree $k+l$. If $K=\alpha \otimes z$ and $L=\beta \otimes v$ with
$z$ and $v$ vector fields, the Fr\"olicher-Nijenhuis bracket can
be characterized by
\begin{eqnarray*}
[\alpha \otimes z&,&\beta \otimes v]_{FN}= \alpha \wedge \beta
\otimes [z,v]+\alpha \wedge L_{z} \beta \otimes v
\\&& \ \ \ -(-1)^{kl}\beta \wedge L_{v} \alpha \otimes z +(-1)^{k}
i_{v}\alpha \wedge d \beta\otimes z \\ && \ \ \  -(-1)^{kl+l}
i_{z}\beta \wedge d \alpha \otimes v,
\end{eqnarray*}
where $[\, ,]$ is the Lie-bracket.

The vector valued curvature $R$ (which has nothing to do with the
Riemann tensor), and its real counterpart $\Omega$ are defined
through the formula
\begin{equation}
2R=-[P, P]_{FN} \qquad \text{or} \qquad 2u^{\beta}
\Omega=-[u^{\mu} \omega, u^{\nu} \omega ]_{FN}.
\end{equation}
After some algebra the previous equation reads
\begin{equation} \label{curvatura2}
\Omega=D \omega \, ,
\end{equation}
that is, the same expression obtained for the usual gauge theory.
This expression is invariant  under multiplication of the
connection and the curvature by the same factor $f(x)$. Previously
that factor was fixed by the requirement $\omega(k)=1$. Here it is
fixed by $\omega(u)=1$.  In this normalization the curvature  is
linked to the vorticity through the relation $\Omega_{\mu
\nu}=-2w_{\mu \nu}$.

Let $A=\p/\p a$, $B=\p/\p b$, be two fields over $S$ of vanishing
Lie bracket, and let $A^{*}$ and $B^{*}$ be their horizontal lift.
Plugging  these fields in Eq.~(\ref{curvatura2}) we find
\begin{equation} \label{sagnac2}
[A^{*}\Delta a, B^{*} \Delta b]^{\mu}u_{\mu}=2w_{\alpha
\beta}\,A^{* \alpha} \Delta a \,B^{* \beta} \Delta b \, .
\end{equation}
The left hand side expresses the holonomy of a closed path $\tau$
of sides $\Delta x^{i}= A^{i} \Delta a$ and $\Delta x^{i}= B^{i}
\Delta b$ in units of proper time. That is, it gives half the
Sagnac effect for the path described. The right hand side stands
for the product $2\vec{\omega} \cdot \vec{\Delta S}$ where
$\vec{\Delta S}$ is the oriented area enclosed by the path $\tau$.
Hence, Eq.~(\ref{sagnac2}) is an infinitesimal, and general
relativistic version of the well known formula of the Sagnac
effect \cite{stedman97}
\begin{equation}
\delta \tau=\frac{4\vec{\omega} \cdot \vec{\Delta S}}{c^2} \, .
\end{equation}
Notice that this equation holds only for infinitesimal paths. In
the case of a stationary frame, there is an integral version given
by Eq.~(\ref{sagnac3}).

In the generalized gauge theory the Bianchi identity reads
\begin{equation}
[P, R]_{FN}=0 \, .
\end{equation}
After some calculations, and using only the horizontality of the
2-form $\Omega$ one arrives at
\begin{equation} \label{bianchi}
D \Omega + a \wedge \Omega=0 \, .
\end{equation}
Hence, the Bianchi identity is a non trivial, and quite
unexpected, differential relation between the vorticity and the
acceleration field. In terms of  the vorticity vector
\begin{equation} \label{vector}
w^{\alpha}=\frac{1}{2}\, \varepsilon^{\alpha \beta \gamma \delta}
u_{\beta} w_{\gamma \delta} \qquad; \qquad w_{\alpha \beta}=
\varepsilon_{\alpha \beta \gamma \delta} u^{\gamma} w^{\delta},
\end{equation}
the Bianchi identity takes the form
\begin{equation}
w^{\nu}_{; \, \nu}+2 w^{\nu} a_{\nu}=0.
\end{equation}
This identity can be checked directly using Eq. (\ref{vorticity})
and Eq. (\ref{vector}). The physical meaning follows recalling
that in flat spacetime and for a non-relativistic fluid described
by a vector field $\vec{v}$ the vorticity vector is
$\vec{\omega}=\frac{1}{2} \nabla \times \vec{v}$. Thus in the
non-relativistic limit $\nabla \cdot \vec{\omega}=0$. The Bianchi
identity is therefore a generalization of this equation to the
relativistic case. It shows that when relativistic effects are
taken into account the acceleration becomes a source for the
vorticity vector. Restoring $c$ one sees that this effect, in the
non-relativistic limit,  is suppressed by a factor $1/c^{2}$.

Table~\ref{table} summarizes the parallelism between simultaneity
and gauge theory.

\section{Back to the usual gauge theory}
One can ask what is the condition that the vector field $u$ should
satisfy in order for the connection to be principal. In other
words, in what circumstance is there a vector field
\begin{equation}
X=\chi(x) u \, ,
\end{equation}
whose one parameter group of diffeomorphisms $\phi^{X}_{t}$ with
$X=\p_{t}$ sends horizontal planes to horizontal planes? If it
exists one can choose
\begin{equation}
\omega_{\mu}=\frac{X_{\mu}}{X \cdot X} \, ,
\end{equation}
and Eqs. (\ref{onormalization}) and (\ref{hinvariance}) become
both satisfied. The invariance of the horizontal planes under the
group of diffeomorphisms $\phi^{X}_{t}$ reads
\begin{equation} \label{hinv2}
\phi^{X}_{t \, *}\, H_{x}=H_{\phi^{X}_{t}(x)} \, .
\end{equation}
Let $\omega$ be a real valued connection ($\text{ker}\, \omega(x)
= H_{x}$). The previous equation becomes
\begin{equation} \label{liex}
L_{X} \omega=f(x) \omega \, ,
\end{equation}
for a suitable function $f: M \to \mathbb{R}$. Let us begin with
$\omega_{\beta}=X_{\beta}$, we find
\begin{equation} \label{xequivalent}
(L_{X} \,g)_{\beta \nu} X^{\nu}=f X_{\beta} \, .
\end{equation}
This equation is equivalent to Eq.~(\ref{hinv2}). It is satisfied,
for instance, if $X$ is a conformal Killing vector field
\begin{equation}
L_{X}g=\frac{X^{\mu}_{; \mu}}{2} \,g \, ,
\end{equation}
hence our previous treatment of simultaneity in stationary frames
can be easily generalized to conformal stationary frames. In both
cases the connection is principal, that is, invariant under the
action of a   structure group. This should have been expected
because the relation of orthogonality follows solely from the
causal structure of spacetime, that is, it is invariant under
conformal transformations of the metric. A conformal Killing
vector field, preserving the causal structure, preserves also the
orthogonality relation and therefore sends horizontal planes to
horizontal planes. Equation (\ref{xequivalent}), however, does not
solve our problem because it is not expressed in terms of the
four-velocity field $u(x)$.

Let $\omega_{\beta}=u_{\beta}$, plugging this equation in
Eq.~(\ref{liex}) we find after some algebra
\begin{equation}
\nabla_{u}u_{\beta}+(\ln \chi)_{\beta}=u_{\beta} \nabla_{u} (\ln
\chi) \, .
\end{equation}
The projection of this equation to the vertical space is
automatically satisfied. It is convenient to introduce the
acceleration 1-form $a_{\beta}=\nabla_{u}u_{\beta}$ and to project
the previous equation to the horizontal plane
\begin{equation} \label{unaltra}
D(\ln \chi)= -a \, .
\end{equation}
The problem is solved if we can establish in what circumstances
this equation has a solution $\ln \chi$ ($\chi^{-1}$ is   the
acceleration potential \cite{ellis71}) over the fiber bundle). The
condition we are seeking is an integrability condition for
Eq.~(\ref{unaltra}). It should be noticed that Eq.~(\ref{unaltra})
gives us the value of $\p_{Y} (\ln \chi)$ only for $Y \in H_{x}$.
In order to find $\p_{u} (\ln \chi)$ we have to use the non
integrability of the distribution of  horizontal planes. If $R=0$,
however, the distribution of horizontal planes becomes integrable
and the spacetime admits a foliation of simultaneity slices. In
that case there is an infinite number of solutions (if any), each
one in correspondence with a different time parameterization of
the simultaneity slices. The equivalence relation "$x \sim y$ if
there exists a piecewise $C^{1}$ horizontal curve which joins $x$
and $y$", divides the spacetime manifold into sets $M_{i}$,
$M=\cup_{i}M_{i}$. In each set the equation (\ref{unaltra}) has a
unique solution $\chi(x)$ up to a constant factor. Indeed if
$\chi'$ is another solution of Eq.~(\ref{unaltra}) then $D\ln(\chi
{\chi'}^{-1})=0$ and hence $\chi {\chi'}^{-1}=c_{i}$  in each set
$M_{i}$ .

Without losing generality we can restrict ourselves to a region
$M_{i}$ where each pair of events  can be joined by a horizontal
curve. We assume that $M_{i}$ is simply connected. Let us consider
the 1-form
\begin{equation} \label{sigma}
\sigma=f(x)u-a \, ,
\end{equation}
where $f(x)$ is a real valued function. With a straightforward
calculation it is easy to show that $\dd \sigma=0$ if and only if
\begin{eqnarray}
Da&=&f\,Du \, , \label{first}\\
L_{u}a&=&f\,a-Df \, . \label{second}
\end{eqnarray}
In this case,  $\sigma=\dd \varphi$ and from Eq.~(\ref{sigma}),
$f=\p_{u} \varphi$ and $D\varphi=-a$. Since $\varphi$ solves
Eq.~(\ref{unaltra}) it is the function $\ln \chi$ we are looking
for. Conversely if $\ln \chi$, solution of (\ref{unaltra}),
exists, it satisfies Eqs.~(\ref{sigma}-\ref{second}) with
$f=\p_{u} \ln \chi$.

The system of equations (\ref{first}) (\ref{second}) is our
necessary and sufficient condition in order to establish whether
our connection is principal. The first equation (\ref{first})
states that the two forms $Da$, $Du$ must be proportional: the
proportionality factor gives  the derivative $f=\p_{u}(\ln \chi)$
that Eq.~(\ref{unaltra}) was unable to determine. The second
equation (\ref{second}) states a condition that can be checked
once we substitute $f(x)$ as given by Eq.~(\ref{first}). Hence, we
can establish whether or not the simultaneity is described by a
principal connection looking only at the physical data $u(x)$.

If $a=0$ the system is solvable with the  solution $\chi(x)=1$.
This result says that, when the congruence of time-like curves is
given by geodesics, the simultaneity can be treated as a gauge
theory over the group of time translations.

Another interesting case, which generalizes the previous one, is
that of a perfect fluid of equation of state $p=p(\rho)$. The
stress-energy tensor is
\begin{equation}
T^{\mu \nu}=(\rho+p)u^{\mu} u^{\nu}-p \,g^{\mu \nu},
\end{equation}
The conservation of 4-momentum $T^{\mu \nu}_{; \nu}$ leads to the
local energy conservation and to the Euler equation
\begin{eqnarray}
\rho_{,\, u}&=& -(\rho+p) \nabla \cdot u \\
(\rho+p)\,a &=& Dp
\end{eqnarray}
One can introduce the functions \cite{ellis71}
\begin{eqnarray}
\chi &=& \exp \left(-\int_{p_{0}}^{p}\frac{\dd p}{\rho+p} \right)\\
R &=& \exp \left(-\int_{\rho_{0}}^{\rho}\frac{\dd
\rho}{3(\rho+p)}\right)=\chi^{-1/3}\left(
\frac{\rho+p}{\rho_{0}+p_{0}}\right)^{-1/3}
\end{eqnarray}
which are related to each other and to the pressure. Using the
relation between the determinants $g=\chi^{2}h$, the
energy-momentum conservation becomes
\begin{eqnarray}
\left( \ln(R^{-3}|h|^{\frac{1}{2}}) \right)_{, u}&=&0 \label{energycons}\\
a+D \ln \chi&=&0
\end{eqnarray}
This last equation says that the flow lines of the perfect fluid
define a frame where the simultaneity is described by a gauge
theory over the group of time translation. The invariance of the
connection $\omega$ under time translations (see Eq.
(\ref{hinvariance}) and Eq. (\ref{connesione2})) implies that the
potential $A_{\sigma}$ does not depend on time. Making use of Eqs.
(\ref{energycons}) (\ref{kaluzam}) (\ref{connesione2}) and
introducing $\bar{h}_{i j}=R^{-2} h_{i j}$,
 the spacetime metric becomes
 \begin{equation*} 
g_{\mu \nu}\dd x^{\mu} \dd x^{\nu}\!= \!\chi(x)^{2}(\dd
t+A_{i}(x^{j})\dd x^{i})^{2}\!+\!R(x)^{2} \bar{h}_{i j}(x^{k}, t)
\dd x^{i} \!  \dd x^{j}
\end{equation*}
where the determinant $\bar{h}$ does not depend on time. The
function  $R(x)$ can be interpreted as the scale factor of the
universe since a volume element scales as $R^{3}$ with time
($\theta=  \nabla \cdot u=3R_{, \, u}/R$). This well known result,
already obtained by Ehlers, Taub and others (see \cite{ellis71}
and references therein) arises here as a natural consequence of a
careful analysis of simultaneity in the simple case of a perfect
fluid with an acceleration potential.

\section{Conclusions}
I have shown that simultaneity  is  described by the mathematics
of gauge theories. The fiber bundle is the spacetime manifold and
the fibers are the trajectories which define the frame.  The
connection is determined by local simultaneous events accordantly
with the Einstein convention, and the horizontal lift follows from
a pointwise Einstein synchronization. The curvature is the angular
velocity of the observers at rest in the frame and the holonomy is
the obstruction against the possibility of applying the Einstein
synchronization in the large. A global simultaneity convention
allows the introduction of a time coordinate on the manifold. A
change in the global simultaneity convention is a gauge
transformation. The gauge formalism suggested the calculation of
the Bianchi identity which turned out to be a differential
relation between the acceleration field and the vorticity field.

 A necessary and sufficient condition on the
vector field $u(x)$ in order for the connection to be principal
has been given. In that case the fiber bundle admits a  structure
group $T_{1}$ of time translations which preserves the connection,
and the usual gauge theory is recovered. Particular cases are
frames of geodesics, frames derived from the flow lines of a
perfect fluid and conformal stationary frames. In this last case
the Sagnac effect depends straightforwardly on the holonomy of the
light path through a generalization of Eq.~(\ref{sagnac3}). Apart
from these simple cases, in order to appreciate the gauge nature
of simultaneity, in the spirit of contemporary physics, the whole
machinery of the Fr\"olicher-Nijenhuis algebra is needed.

The gauge picture of simultaneity is expected to be useful in the
search for non trivial solutions of the Einstein equations.
Moreover, once generalized to higher dimensions, the mathematics
involved can prove to be useful in modern Kaluza-Klein theories,
since there, the cylinder condition is dropped.

\section*{Acknowledgments} I wish to thank B. Carazza, M. Modugno and C. Destri for
support, criticisms and useful discussions. This work was
partially supported by INFN.\\


\end{document}